# Rare-earth ions incorporation into Lu$_2$Si$_2$O$_7$ scintillator crystals: Electron paramagnetic resonance and luminescence study

M. Buryi[1], V. Laguta[1], V. Babin[1], O. Laguta[2], M.G. Brik[3,4,5], M. Nikl[1]

[1]*Institue of physics CAS in Prague, Cukrovarnická 10/112, Prague, Czech Republic*
[2]*Brno University of Technology, Central European Institute of Technology, Purkyňova 656/123 61200 Brno, Czech Republic*
[3]*CQUPT-BUL Innovation Institute & College of Sciences, Chongqing University of Posts and Telecommunications, Chongqing 400065, People's Republic of China*
[4] *Institute of Physics, University of Tartu, W. Ostwald1 1, Tartu 50411, Estonia*
[5] *Institute of Physics, Jan Długosz University, Armii Krajowej 13/15, PL-42200 Częstochowa, Poland*

**Abstract**

The present work reports results of the electron paramagnetic resonance (EPR), optical absorption, radio- and photoluminescence (RL and PL) complex study of the Lu$_2$Si$_2$O$_7$:Pr and Lu$_2$Si$_2$O$_7$:Ce pyrosilicate crystals. In both crystals, the EPR spectra demonstrate the presence of characteristic signals originating from the Yb$^{3+}$, Er$^{3+}$, Nd$^{3+}$, Dy$^{3+}$, Gd$^{3+}$ and V$^{3+}$ ions existing in the material as uncontrolled impurities. The corresponding spectra (except for Gd$^{3+}$) have been analysed in detail and **g**- and hyperfine tensors are determined for all these ions in the lutetium pyrosilicate for the first time. Optical absorption, RL and PL measurements in the Lu$_2$Si$_2$O$_7$:Pr crystal demonstrated presence of only characteristic Pr$^{3+}$ transitions. In addition, the Pr$^{3+}$- Pr$^{3+}$ energy transfer was observed and confirmed experimentally.

## 1. Introduction

Scintillation materials presently are widely used in various technical fields, the most important of which are the high energy physics, medical imaging, industrial controls, homeland security and others [1]. They must be very efficient in conversion of high energy photons or particles into easily detectable ultraviolet or visible light. Unstoppable search for improvement of already existing and development of new scintillating materials results in the incredibly increased number of the appropriate hosts and dopants. One of those is the Ce$^{3+}$-doped lutetium pyrosilicate, Lu$_2$Si$_2$O$_7$ (LPS), a perspective scintillator for application in the positron emission tomography (PET) [2]. The Ce$^{3+}$ ion was expected to occupy in this lattice the regular lutetium site what was proved by electron paramagnetic resonance (EPR) investigation of the LPS:Ce crystal [3]. The LPS:Ce single crystals grown by the melting zone technique exhibit relatively high light yield (LY) about 20000-27000 ph/MeV along with about 37 ns cerium scintillation decay time and no detectable afterglow [4,5]. Another activator dopant used in the LPS-based scintillators is Pr$^{3+}$. It appeared in the crosshair of scientific interest as a fast scintillator from 1990's [6,7]. The Pr$^{3+}$ demonstrated faster 5d–4f emission than the Ce$^{3+}$ in the microcrystalline LPS prepared by sol-gel method, with the decay times of 16 ns and 37 ns, respectively [8]. Therefore, the Pr-doped lutetium pyrosilicate attracts considerable attention as an ultrafast scintillator.

It is commonly known that the higher light yield and the faster scintillation decay provide the better resolution and larger signal-to-noise ratio in the imaging techniques, either in medicine or in other fields of human activity. Moreover, the constant light yield, independent of the absorbed dose of ionizing radiation, is required especially in medical imaging. Its instability (increase or decrease upon the dose) causes doubling of the observed image [9]. Some scintillators suffer the dose dependence of the light yield, e.g., thallium doped caesium iodide [10], the material widely used in X-ray imaging [11,12]. Praseodymium-doped lutetium pyrosilicate was also investigated on the subject [13]. The radioluminescence (RL) efficiency was shown to increase because of the effective charge traps filling and further depletion resulting in the thermoluminescence (TL) glow peaks at 460 and 515 K creating concurrence to the $Pr^{3+}$ emission centers in free carrier trapping under the progressive irradiation. The deliberation of the charge recovers afterwards the initially lowest RL efficiency. The observed an a-thermal afterglow made evidence for the spatial correlation between traps and $Pr^{3+}$ ions. However, the precise description of the physical mechanisms lying behind the phenomenon is often unclear. It is obviously related to the presence of defect (trapping states) in the host bandgap. Therefore, revealing the traps and competitive recombination centers and study their origin is of great importance for their suppression, neutralization or simply prevention during the materials synthesis.

As a rule, the deep traps are caused by the impurity ions of transition series introduced into the growing ingot from crucible, either from the crucible material itself or leftovers of the compounds used for previous substances fabrication. The purity of residual reagents is also of very high priority, since many of the impurities found in the starting materials can appear afterwards in the resulting scintillator host. As an example, the $LuAlO_3$:Ce thin films [14], $La_3Ta_{0.5}Ga_{5.5}O_{14}$ single crystals [15], sulphides [16], tungstates single crystals [17,18] fabrication can be mentioned. Moreover, the Lu and Y oxides may contain the whole range of lanthanides, especially $Yb^{3+}$ as it was shown in e.g., $Lu_3Al_5O_{12}$:Ce [19]. Aluminum-based oxides demonstrate also the presence of $Cr^{3+}$, $Fe^{2+/3+}$, $Mo^{n+}$ (n = 1-6) [20,21], $V^{n+}$ (n = 2-5) [22, 23], commonly known natural accidental impurity ions found in corundum. Silicon-based oxides can contain unintentionally $Ge^{4+}$ (see e. g., [24]) and P [25,26]. The impurity ions beside of the charge trapping sites can also provide other, unexpected ways of energy transfer.

To the best of our knowledge, very little work is done towards the study of impurities in LPS crystals [27,28] and even fewer papers report the results of the traps and impurities origin in this material. In this connection, the EPR study of $Ce^{3+}$ ions [3] and traps related to $Ir^{3+/4+}$ ions [29] can be mentioned here. Therefore, the present research is mainly focused on complete EPR study of the Pr- and Ce-doped lutetium pyrosilicates aiming at collection of exhausted information concerning the type of uncontrolled impurities, their localization in lattice and charge state, magnetic and optical properties, and possible role in energy transfer processes induced by high-energy irradiation.

## 2. Experimental

The Lu$_2$Si$_2$O$_7$:Pr (LPS:Pr) and Lu$_2$Si$_2$O$_7$:Ce (LPS:Ce) crystals were grown from melt in iridium crucibles by Czochralski method in atmosphere of high pure nitrogen. The initial praseodymium and cerium concentrations in the melt were 0.5 and 0.3 at. %, respectively, with respect to the total rare earth sites. LPS crystallize in the monoclinic structure, space group C2/m [30]. In this lattice there is a single crystallographic site for Lu$^{3+}$ ions with six oxygen neighbors forming a distorted octahedron with C$_2$ symmetry. It is expected that the trivalent rare earth (RE$^{3+}$) impurity ions will substitute namely for the Lu$^{3+}$ ions due to close ionic radii as compared to the small Si$^{4+}$ ion and the same valence state.

Radioluminescence measurements were performed at room temperature using a Seifert X-ray tube with tungsten target operated at 40 kV; the RL emission was detected by a Horiba Jobin-Yvon setup and detected by a TBX-04 photomultiplier tube (IBH) in the 200-800 nm spectral range. The spectral slitwidth of the emission monochromator was 8 nm.

Electron paramagnetic resonance (EPR) measurements were performed in the X-band (9.4 GHz) with a commercial Bruker X/Q-band E580 FT/CW ELEXSYS and EMX spectrometers at temperatures 10-30 K. For EPR measurements, the crystals were cut in three orthogonal planes (*a∗b*), (*bc*), and (*a∗c*). The *a*\* axis was deflected from the crystallographic axis *a* by an angle of 12° in order to satisfy the orthogonality between crystal planes that was necessary for determination of g factors from EPR spectra.

## 3. Results and discussion

### *3.1. EPR spectra of RE ions*

The LPS crystals were intently doped with Pr or Ce only, whereas the measured EPR spectra show the presence of other rare-earth ions in the paramagnetic state, namely Yb$^{3+}$, Dy$^{3+}$, Er$^{3+}$, and Nd$^{3+}$. The V$^{3+}$ EPR signals were also detected. All these paramagnetic ions penetrate into the crystal from raw materials as uncontrolled impurities. Corresponding spectral lines of all these ions are indicated in the spectrum in Fig. 1 recorded at 18 K. Excepting the Ce$^{3+}$, all other ions show characteristic hyperfine structure in the spectrum originating from isotopes with non-zero nuclear magnetic moments. Therefore, all these ions were easily identified. Separately, EPR spectra of the Yb$^{3+}$ Dy$^{3+}$, Er$^{3+}$, Nd$^{3+}$, and V$^{3+}$ ions are presented in Fig. 2. In order to show all details of the hyperfine (HF) structure in their spectrum, each of these spectra was measured at temperature and crystal orientation which are optimal for each of the ions. The HF patterns were simulated taking into account natural abundances of isotopes and their nuclear magnetic moments by using the Bruker WinEPR SimFonia program.

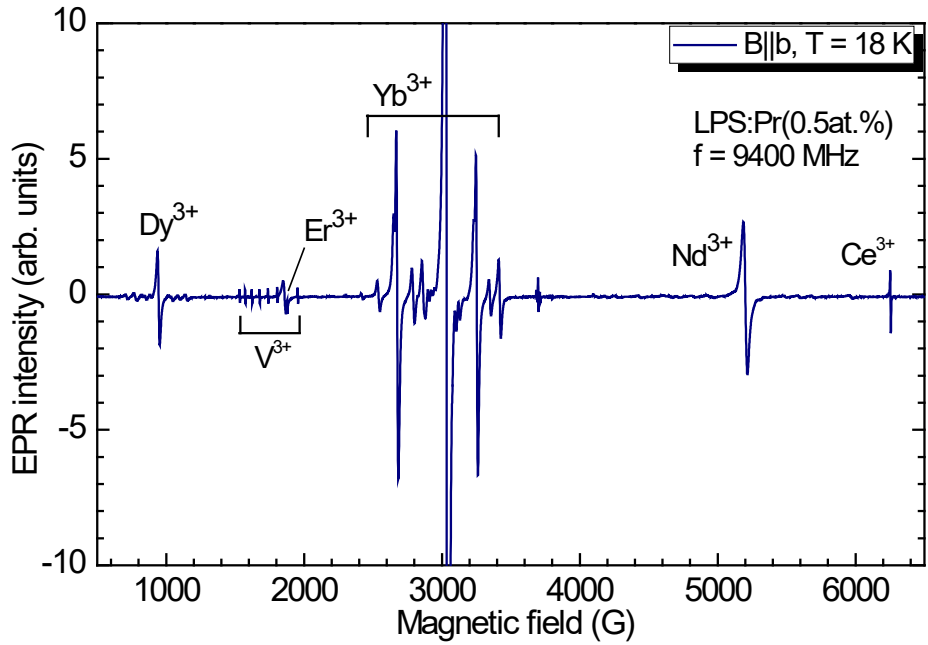

**Fig. 1.** EPR spectrum measured in LPS:Pr single crystal at the orientation of external magnetic field **B** ∥ **b** at 18 K. Spectral lines of identified ions are indicated.

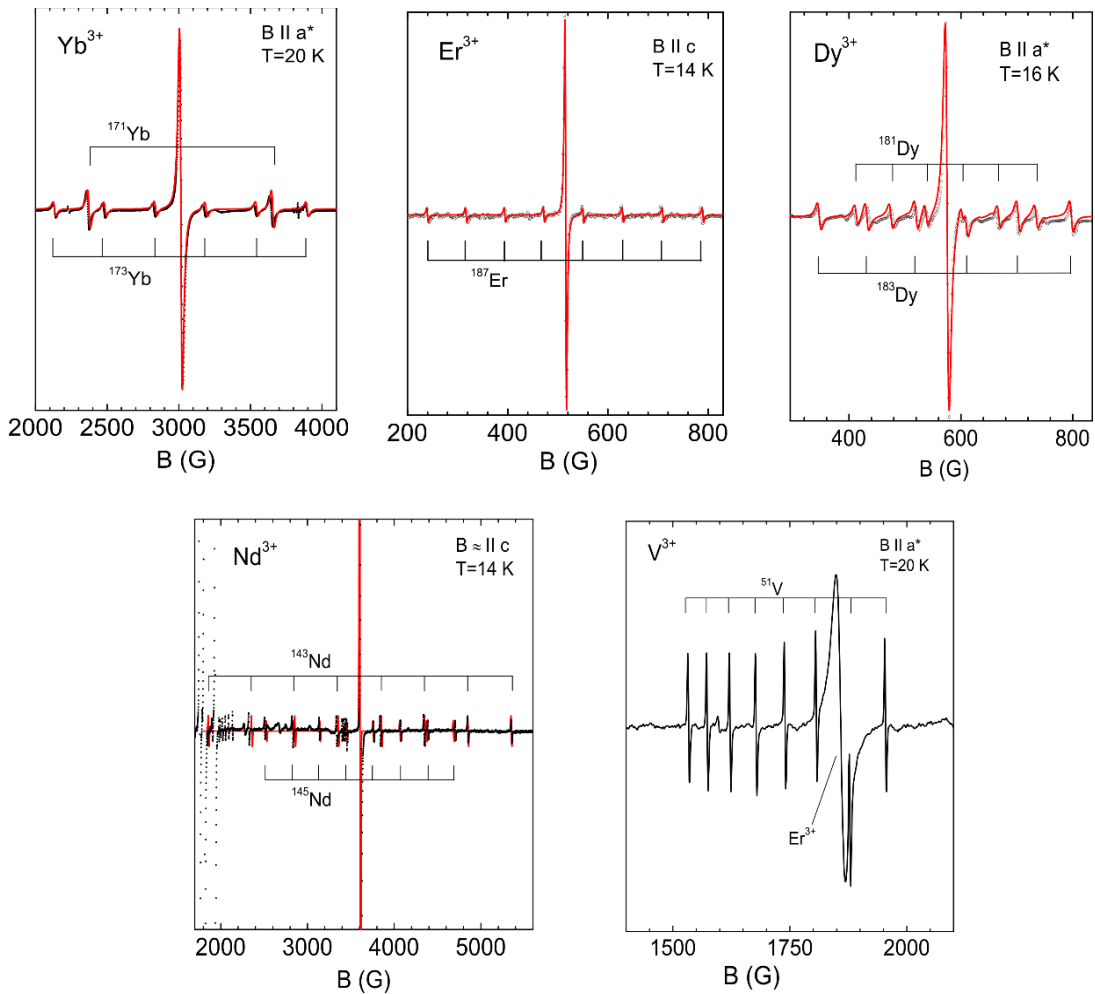

**Fig. 2.** $Yb^{3+}$, $Er^{3+}$, $Dy^{3+}$, $Nd^{3+}$, and $V^{3+}$ spectra in LPS:Pr showing HF structure from isotopes with non-zero nuclear magnetic moments and their numerical simulation (red solid lines). The simulation was done by using Bruker WinEPR SimFonia program. $V^{3+}$ spectrum contains eight HF components of equal intensity due to 100% natural abundance of the $^{51}V$ isotope, which has the nuclear spin $I = 7/2$.

In order to determine spectroscopic parameters of the $Yb^{3+}$, $Er^{3+}$, $Nd^{3+}$, $Dy^{3+}$, and $Ce^{3+}$ ions, three sets of angular dependencies of the centers of gravity of each of the RE ions spectrum (stronger central lines, which correspond to isotopes with zero nuclear magnetic moment) in three orthogonal planes, (bc), (a*b) and (a*c), were measured and constructed in Fig. 3.

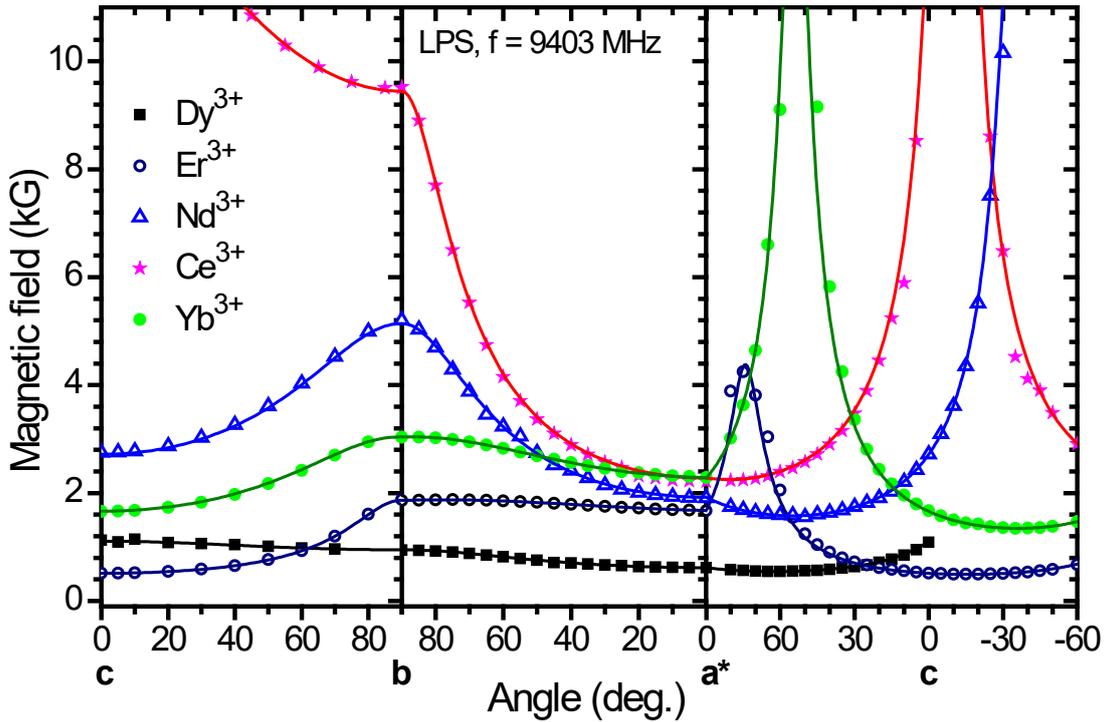

**Fig. 3.** Angular dependencies of the centers of gravity of the resonances originating from $Yb^{3+}$, $Dy^{3+}$, $Er^{3+}$, $Nd^{3+}$ and $Ce^{3+}$ (since the Ce nucleus occurs only as the isotopes with zero nuclear magnetic moment, no HFS could be observed and thus the resonance line is a singlet). Dots represent the experimental data whereas the colour solid lines are the calculated ones.

The following spin Hamiltonian was used for angular fitting of the experimental resonance lines positions (solid lines in Fig. 3):

$$\hat{H} = \beta_e \hat{S} g H, \qquad (1)$$

here $\beta_e$ is the Bohr magneton, **g** is a $g$ tensor (variation parameter in the fitting procedure), $\hat{S}$ and **H** are electron spin operator and magnetic field, respectively. Determined principal values and axes of the **g** tensors are listed in Table 1.

Table 1. Spin-Hamiltonian parameters ($g$ and HF tensors principal values and Euler angles of the principal axes) of RE ions in LPS crystal. The error margins of the $g$ and HF tensors components are 0.005 and 20-50 MHz, respectively. The error margin of the Euler angles is approximately 5º. The HF components are given only for the following isotopes: $^{171}$Yb, $^{163}$Dy, $^{167}$Er, $^{143}$Nd. Those for the $^{173}$Yb, $^{161}$Dy, $^{145}$Nd can be easy calculated using nuclear magnetic moments and data in the table.

| Ion | $g_1$ | $g_2$ | $g_3$ | $\alpha$, º | $\beta$, º | $\gamma$, º | $A_1$ (MHz) | $A_2$ (MHz) | $A_3$ (MHz) |
|---|---|---|---|---|---|---|---|---|---|
| $Ce^{3+}$ | 2.980 | 0.710 | 0.100 | 0 | 80 | 0 | - | - | - |
| $Dy^{3+}$ | 2.097 | 7.071 | 12.263 | 0 | 60 | 0 | 150 | 850 | 1470 |
| $Er^{3+}$ | 1.514 | 3.606 | 13.567 | 0 | -15 | 4 | 220 | 370 | 1450 |
| $Nd^{3+}$ | 0.478 (0.565)* | 1.304 | 4.247 | 0 | 55 | 0 | 220 | 370 | 1350 |
| $Yb^{3+}$ | 0.125 | 2.211 | 4.983 | 0 | -35 | 3 | 1700 | 1700 | 3900 |

*) $g_1$ = 0.565 was determined from powder spectrum.

The values determined for the $Ce^{3+}$ ions are in perfect agreement with those previously determined for the $Ce^{3+}$ in $Lu_2Si_2O_7$ in [3]. **g** tensor values determined for rest of the RE ions are of the same order of magnitude as reported in other materials [31].

The $Dy^{3+}$, $Er^{3+}$, $Nd^{3+}$ resonances are approximately 90 times weaker in intensity than the $Yb^{3+}$ ones (Fig. 1). The corresponding HF structures of these ions have even several times lower intensity than the central lines and, moreover, are overlapped too strongly with other resonances at most of the magnetic field orientations with respect to the crystal axes. Thus, it was impossible to draw the angular dependences of the $^{161,163}$Dy, $^{143,145}$Nd, and $^{187}$Er HF components. It could be done only for the $^{171}$Yb isotopes having nuclear spin $I$ = 1/2 and 14.4% abundance (only the doublet of lines originating from the $^{171}$Yb HFS was clearly resolved in angular dependencies among other HF structures). Moreover, the HF structure from the $^{173}$Yb isotope, having $I$ = 5/2 and 16.2% abundance was also badly resolved at most crystal orientations (its one HF line is approximately 3 times lower in intensity than that of the $^{171}$Yb HF structure, which contains also numerous forbidden transitions due to large quadrupole interaction). Angular dependencies of the $^{171}$Yb HF resonances are shown in Fig. 4.

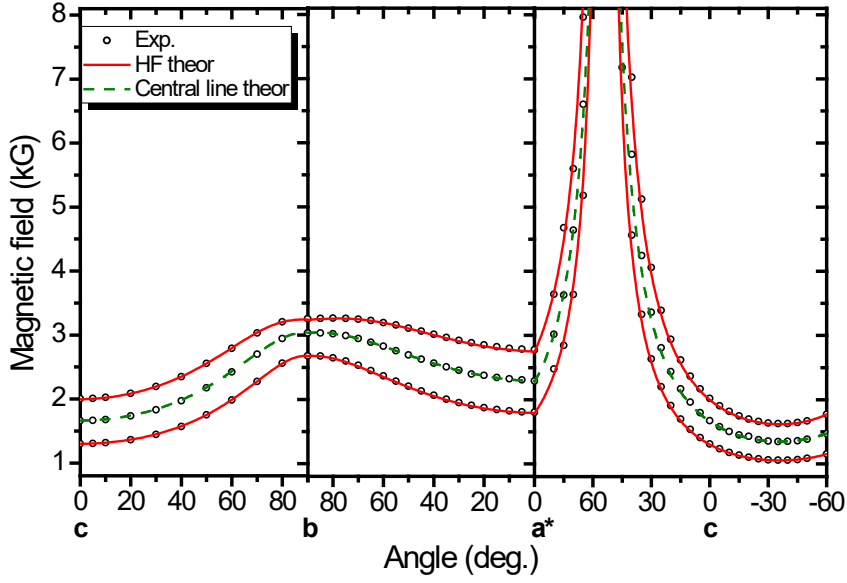

**Fig. 4.** Angular dependencies of the $^{171}$Yb HF resonance fields together with the central line resonances. The experimental data are shown by the dots, red solid lines are the fitting curves of the HF components whereas the green dashed lines fit the central line positions.

The HF structure angular dependencies in Fig. 4 were fitted by the following spin Hamiltonian:

$$\hat{H} = \beta_e \hat{\mathbf{S}} \mathbf{g} \mathbf{H} + \hat{\mathbf{S}} \mathbf{A}(\text{Yb}) \hat{\mathbf{I}}(\text{Yb}), \qquad (2)$$

where **A(Yb)** and **I(Yb)** are the HF tensor and nuclear spin operator for the $^{171}$Yb isotope. It has been found that the HF tensor principal axes coincide with those of the **g** tensor, given by the same Euler angles. The $^{171}$**A** tensor principal values are listed in Table 1. The HF tensor for the second $^{173}$Yb isotope can be easily calculated by using nuclear magnetic moments of two Y isotopes.

To determine the values of HF tensors of the rest of rare-earth ions and their isotopes, the spectra measured at the magnetic field along **a***, **b** and **c** crystal directions have been fitted with the calculated ones taking into account the corresponding **g** tensors from Table 1. Some of these spectra with the fitted HF structures are presented in Fig. 2. Similarly to the Yb$^{3+}$ ions, the HF tensor principle axes orientations were taken from the **g** tensor axes. In reality, a small deviation between the two frames of these axes can exist. However, it can hardly be determined from our data due to influence of quadrupole effects, which also shift spectral lines (note that along crystal axes quadrupole effects are small). The clarification of these effects needs a separate study.

The HF parameters obtained from the fits of HF structures are listed in Table 1. The determined HF constants are rather large demonstrating common property of the HF interaction of the rare-earth ions [32].

To check whether the **g** tensor principal values determined from the fit of angular dependencies are physically based, single crystals with increased Ce$^{3+}$ and Yb$^{3+}$ contents, grinded into powder, were used. The EPR spectra in powders were measured using the electron spin echo (ESE) detected EPR. The advantage of this method is measurement of pure absorption signal (not the derivative of EPR signal as in case of continuous wave (CW) EPR) and thus the whole spectral intensity is preserved in the ESE detected EPR spectra. Moreover, in the ESE detected spectrum, signals from different ions can be better separated from each other because different ions are characterised by different spin-lattice and spin-spin relaxation times which substantially influence signal intensity.

Corresponding ESE detected spectra are shown in Fig. 5a. In particular, EPR spectrum measured in LPS with a large concentration of the Yb ions shows two well-seen singularities at $g$ factors 4.98 and 2.21, which correspond to two g factor principal values. They are in perfect agreement with similar values determined from the fit of resonance field angular dependencies. The resonance magnetic field value, which should correspond to the third g factor, is out of the allowed magnetic field range (beyond the magnet abilities). In case of the Ce doped LPS, there is a superposition of two spectra: $Ce^{3+}$ and $Yb^{3+}$. The latter is always presented in large enough concentration even in Yb undoped crystals. Only one $Ce^{3+}$ peak is well seen at $g = 0.71$. This value is also in perfect agreement with data in Table 1. On the other hand, it was possible to derive more information from the CW EPR spectrum in powder shown in Fig. 5b, including identification of some resonances belonging to other RE ions after comparison with the spectra in Fig. 5a. Especially, there are visible peaks at $g(Ce) = 2.98$ and $g(Nd) = 4.25$, which are in agreement with the corresponding values in Table 1. The $g = 0.565$ value is close to the determined $g_1 = 0.478$ of the $Nd^{3+}$ ion and most probably namely the $g = 0.565$ is the correct value for this ion.

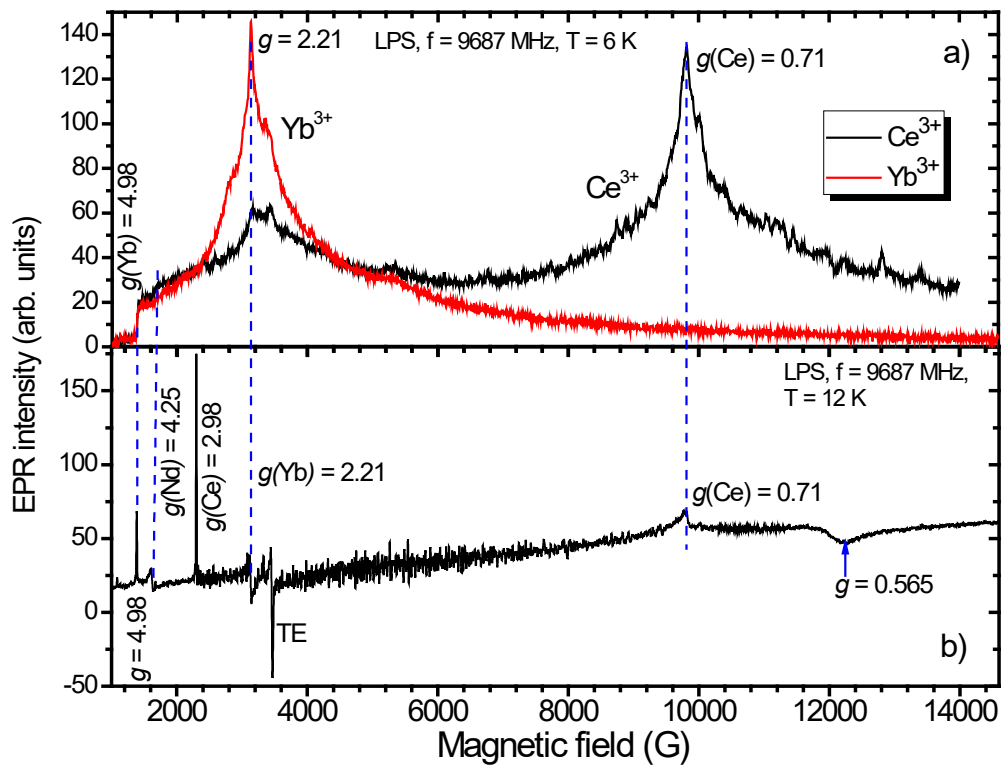

**Fig. 5.** (a) ESE detected $Yb^{3+}$ and $Ce^{3+}$ spectra in LPS powder. (b) CW EPR spectrum measured in the same LPS powder. "TE" means a signal from trapped electron center.

As for the resonances of vanadium (Figs. 1 and 2), it was deduced that it is in the +3 charge state ($3d^2$ ion, electron spin $S = 1$). Its 8 resonance lines of HF structure due to $^{51}$V nucleus (nuclear spin $I = 7/2$, 100% abundance) appear centered at magnetic field values never passing over 2200 G in the X MW band. The corresponding $g$ factor is thus not smaller than approximately 3.05. This is completely inconsistent with the $V^{4+}$ ($3d^1$ ion) exhibiting $g$ factor values less than 2 (see e.g., very recent investigation of the vanadium ion charge states by EPR in $Y_3Al_5O_{12}$ [33]). Therefore, this rather points out large zero field splitting (ZFS) constant of this $d^2$ ion, greater than the microwave quantum used in the X band EPR measurements.

The $d^2$ configuration with the electron spin $S = 1$ has two spin allowed transitions -1 ↔ 0 and 0 ↔ +1. There are also possible -1 ↔ +1 electron spin forbidden transition and double quanta absorption [34]. Since, the observed $V^{3+}$ spectral lines (Figs. 1 and 2) are quite narrow as for the $d^2$ ion they must belong to the double quantum absorption. Determination of the spin Hamiltonian parameters in this case are problematic. However, the situation is much simpler in measurements at much higher microwave frequencies than the 9.4 GHz. We have performed EPR measurements at 210 GHz by using the spectrometer described in [35]. Because the crystal was not doped with vanadium, the $V^{3+}$ EPR signal is quite weak. It was possible to detect only one of the allowed transitions. Its spectrum is shown in Fig. 6. It corresponds to the -1 ↔ 0 allowed transition. The $^{51}$V HF splitting is $82 \times 10^{-4}$ cm$^{-1}$. Such splitting is within the range of values usually measured for $V^{3+}$ ion in a tetrahedral oxygen environment [31]. Assuming that $g$ factor for the tetrahedral $V^{3+}$ is in the range 1.92 – 1.96, one can estimate the zero-field splitting value for $V^{3+}$ ion in LPS lattice. It is in the 0.93 – 0.98 cm$^{-1}$ range. This value also agrees with the tetrahedral position of the $V^{3+}$ ions [31,33]. It means that this impurity ion substitutes for $Si^{4+}$. On the other hand, all $RE^{3+}$ ions in this host are expected to occupy lutetium site because of close ionic radii values [36] and close chemical properties with Lu.

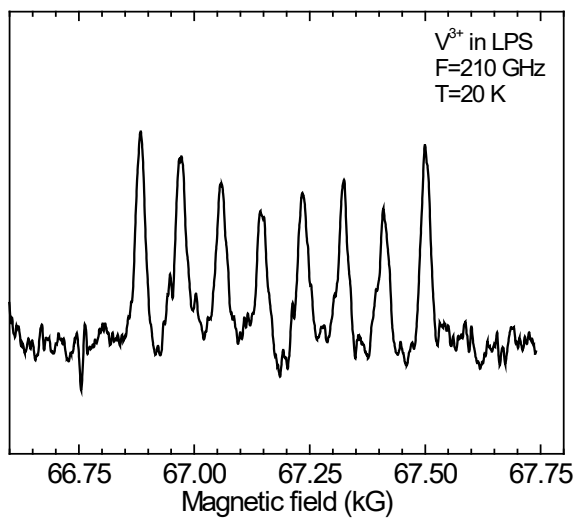

**Fig. 6.** $V^{3+}$ EPR spectrum in LPS crystal measured at B || c and MW frequency 210 GHz.

*3.2. Optical characterization*

One of the studied crystal was doped with 0.5% Pr. The $Pr^{3+}$ EPR spectrum was not visible in our measurements even at 210-320 GHz as expected EPR transitions are located in the infrared region. Therefore, optical absorption spectrum has been measured in this crystal (Fig. 7). Only the transitions characteristic for the $Pr^{3+}$ were observed in according with already published data [37].

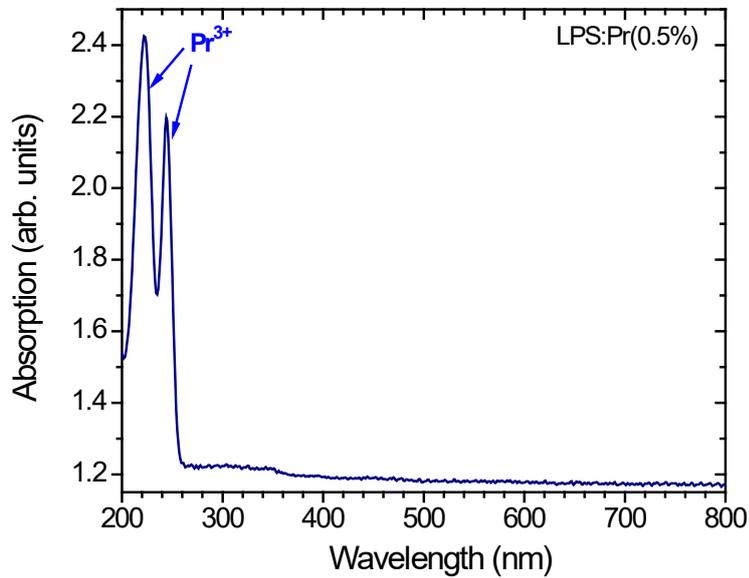

**Fig. 7.** Optical absorption spectrum measured in the LPS:0.5%Pr at 296 K.

RL spectrum of the same crystal is shown in Fig. 8. As in the absorption spectrum, only the transitions originating from the $Pr^{3+}$ 5d-4f and 4f-4f transitions were observed and resolved according to the data published in [27,38]. No traces of other ions reported above were detected at all.

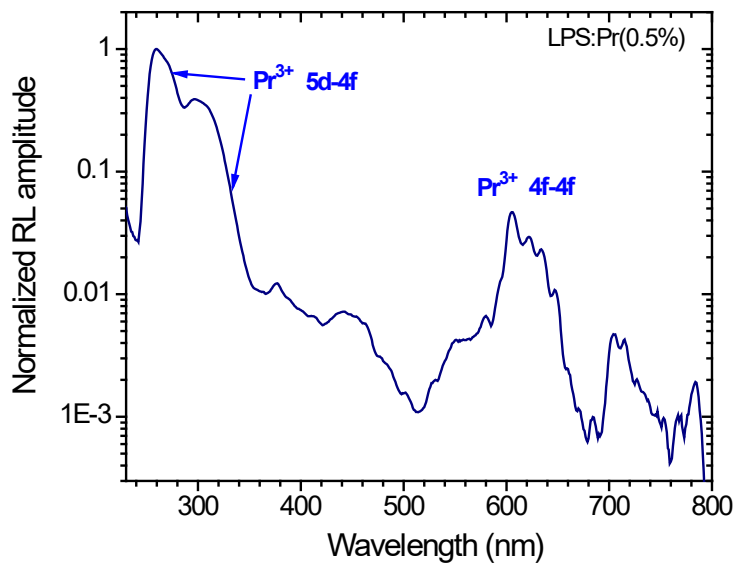

**Fig. 8.** RL spectrum measured in the LPS:0.5%Pr. Regions of the $Pr^{3+}$ 5d-4f and 4f-4f transitions are indicated.

To gain a better insight into the luminescence peculiarities and to study possible energy transfer processes in the lutetium pyrosilicate, the photoluminescence emission (PL) and excitation (PLE) spectra were measured. The PLE spectra were measured for the 265 nm (5d-4f transitions, Fig. 9) and 606 nm (4f-4f transitions, Fig. 10), the strongest $Pr^{3+}$ RL (Fig. 8) and PL (Fig. 9) emissions.

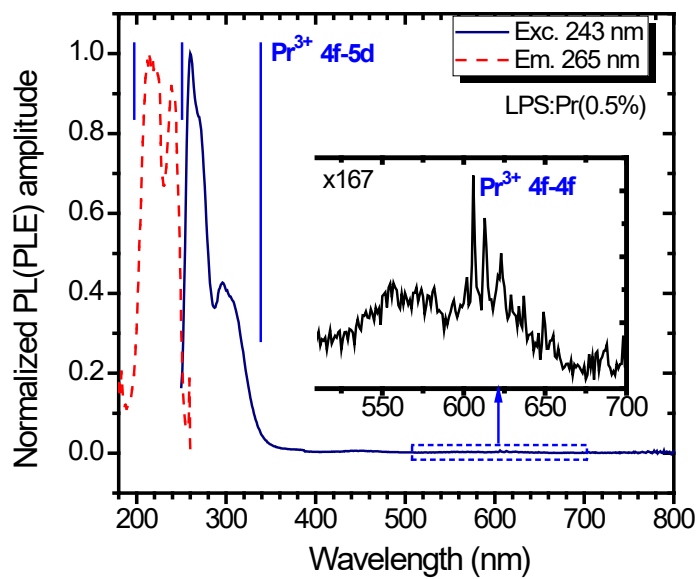

**Fig. 9.** PLE and PL spectra measured in the LPS:0.5%Pr. Corresponding wavelengths are given in a legend. The detected 4f-4f $Pr^{3+}$ transitions are demonstrated as zoomed in an inset.

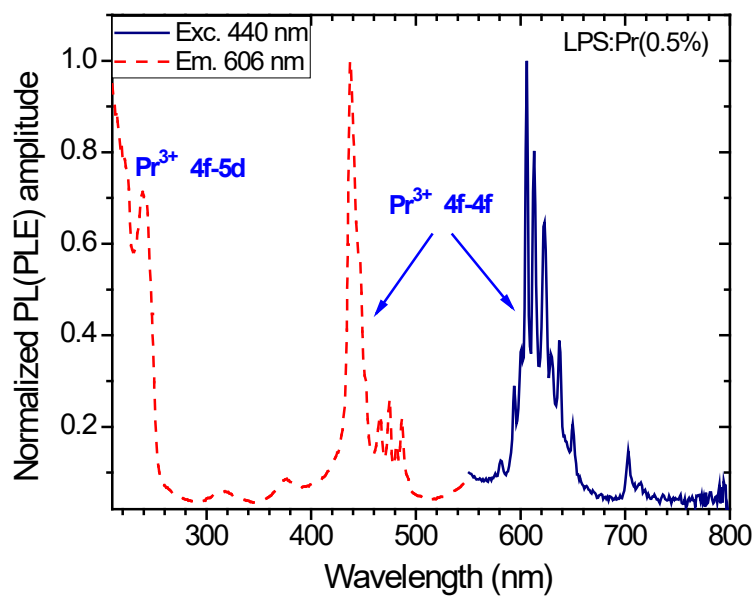

**Fig. 10.** PLE and PL spectra measured in the LPS:0.5%Pr. Corresponding wavelengths are given in a legend.

Remarkably, the PLE spectrum measured for the 606 nm 4f-4f transition contains also the 4f-5d one of comparable intensity. This might indicate also energy transfer (ET) between different $Pr^{3+}$ centers. Taking into account the initial 0.5 at.% praseodymium ions doping level, the total probability of finding two neighbouring $Pr^{3+}$-$Pr^{3+}$ ions can reach about 5-8%. The estimation has been done basing on the similar expectations obtained for the $Ce^{3+}$- $Ce^{3+}$ coupled ions in $YAlO_3$ [39]. In order to clarify more this point, the luminescence kinetics have been measured. The corresponding decay curves are shown in Fig. 11 for the 300 nm (5d-4f) and 606 nm (4f-4f) emission. The 256 nm (4f-5d) and 437 nm (4f-4f) excitation wavelengths, respectively, were used. The 300 nm and 256 nm excitations in the case of the d↔f transitions were chosen on the edges of the PLE and PL spectra in order to compensate small Stokes shift (Fig. 9). The experimental data in the both cases were fitted by the commonly known single exponential decay function exhibiting the 20 ± 2 ns (300 nm) and 210 ± 5 us (606 nm) decay times, respectively. However, the fit in Fig. 11b is not ideally single exponential. There one may expect very small second component. The respective place is indicated in Fig. 11b in an inset in this figure. Nothing similar was observed in Fig. 11a. Therefore, above-mentioned efficient $Pr^{3+}$-$Pr^{3+}$ ET can really contribute to the excited states kinetics processes.

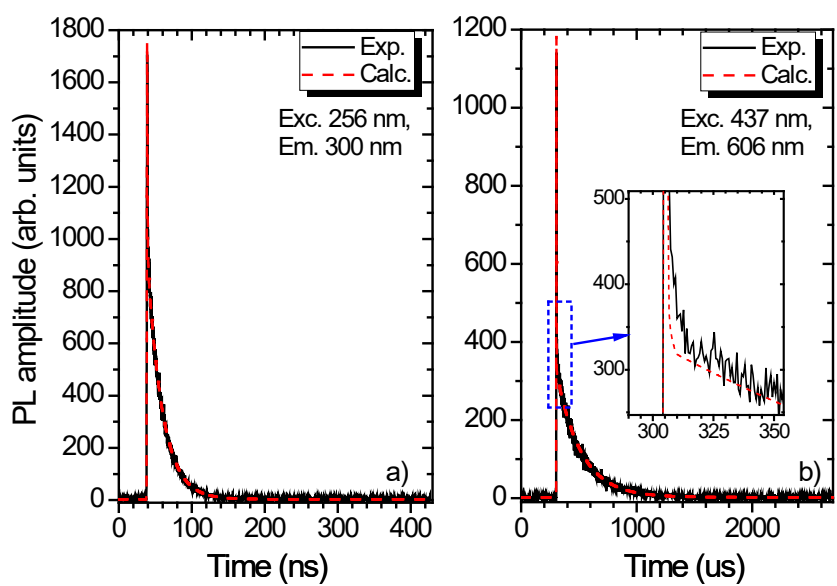

**Fig. 11.** Decay curves for (A) excitation at 256 nm, 300 nm emission; (B) excitation at 437 nm, 606 nm emission. Experimental (black solid lines) and fitting (single exponential, red solid lines) data are distinguished by color. An inset in the panel B demonstrates slight disagreement between the experiment and fit. The place may stress the possible presence of the second component.

No other bands, which might be connected with other RE ions, were observed within 200-800 nm range (restricted to the photodetector sensitivity). This agrees well with low concentrations of the uncontrolled RE impurity ions discussed in subsection 3.1. The $Yb^{3+}$ signal strongly dominates in the EPR spectra (Fig. 1). However, its optical transitions cannot be observed in the mentioned range since they appear in the infrared region [40].

**Conclusions**

Yb$^{3+}$, Ce$^{3+}$, Er$^{3+}$, Nd$^{3+}$, Dy$^{3+}$, and V$^{3+}$ ions as uncontrolled impurities were revealed and studied in detail in LPS:0.5%Pr scintillation crystals by measuring their EPR spectra. The same RE ions are presented in Ce-doped LPS crystals. Corresponding **g** and hyperfine tensors were determined for the first time for the all these RE ions by analysis of single crystal and powder spectra. The Yb$^{3+}$ signal strongly prevails in the EPR spectra. The RE ions occupy lutetium site whereas the vanadium ions presumably substitute for the silicon ions. According to optical absorption, RL and PL measurements, only Pr$^{3+}$ spectra were detected in LPS:Pr. No traces of other ions were observed at all because of their low concentration. The Yb$^{3+}$ optical transitions occur in the infrared region so they were not measured presently in the 200-800 nm experimental region. The Pr$^{3+}$- Pr$^{3+}$ energy transfer was confirmed by the analysis of the PLE spectra measurements for the 606 nm 4f-4f transition where the 4f-5d transition was detected as well of the comparable intensity with the 4f-4f one.

**Acknowledgements**

The authors gratefully acknowledge the financial support of the Czech Science Foundation (project No. 17-09933S), the Ministry of Education, Youth and Sports of Czech Republic under project CZ.02.1.01/0.0/0.0/16_013/0001406. M.G.B. acknowledges the support from the National Recruitment Program of Highend Foreign Experts (Grants No. GDT20185200479 and No. GDW20145200225), Programme for the Foreign Experts (Grant No. W2017011) and Wenfeng High-end Talents Project (Grant No. W2016-01) offered by Chongqing University of Posts and Telecommunications (CQUPT), Estonian Research Council Grant No. PUT PRG111, European Regional Development Fund (TK141), and NCN Project No. 2018/31/B/ST4/00924. We are thankful to Prof. Guohao Ren from Shanghai Institute of Ceramics, CAS, for providing us the crystals for this study.

**References**

[1] M. Nikl, A. Yoshikawa, Recent R&D trends in inorganic single-crystal scintillator materials for radiation detection, Adv. Opt. Mater. **3** (2015) 463.

[2] L. Pidol, A. Kahn-Harari, B. Viana, E. Virey, B. Ferrand, P. Dorenbos, J.T.M. De Haas, C.W.E. Van Eijk, High efficiency of lutetium silicate scintillators, Ce-doped LPS, and LYSO crystals. IEEE Trans. Nucl. Sci. **51** (2004) 1084–1087.

[3] L. Pidol, O. Guillot-Noël, A. Kahn-Harari, B. Viana, D. Pelenc, D. Gourier, EPR study of Ce$^{3+}$ ions in lutetium silicate scintillators Lu2Si2O7 and Lu$_2$SiO$_5$, J. Phys. Chem. Solids **67** (2006) 643–650.

[4] D. Pauwels, N. Le Masson, B. Viana, A. Kahn-Harari, E.V.D. van Loef, P. Dorenbos, and C.W.E. van Eijk, A Novel Inorganic Scintillator: Lu$_2$Si$_2$O$_7$:Ce$^{3+}$ (LPS), IEEE Trans. Nucl. Sci. **47** (2004) 1787-1790.

[5] L. Pidol, A. Kahn-Harari, B. Viana, E. Virey, B. Ferrand, P. Dorenbos, J.T.M. de Haas, C.W.E. van Eijk, High Efficiency of Lutetium Silicate Scintillators, Ce-Doped LPS, and LYSO Crystals, IEEE Trans. Nucl. Sci. 51 (2004) 1084-1087.

[6] H. Suzuki, T.A. Tombrello, C.L. Melcher, J.S. Schweizer, UV and gamma-ray excited luminescence of cerium-doped rare-earth oxyorthosilicates, Nucl. Instrum. Methods Phys. Res. A 320 (1992) 263.

[7] C.L. Melcher, R.A. Manente, C.A. Peterson, J.S. Schweizer, Czochralski growth of rare earth oxyorthosilicate single crystals. J. Cryst. Growth 128 (1993) 1001.

[8] M. Nikl, G. Ren, D. Ding, E. Mihokova, V. Jary, H. Feng, Luminescence and scintillation kinetics of the $Pr^{3+}$ doped $Lu_2Si_2O_7$ single crystal, Chemical Physics Letters 493 (2010) 72–75.

[9] A.R. Cowen, A.G. Davies, M.U. Sivananthan, The Design and Imaging Characteristics of Dynamic, Solid-State, Flat-Panel X-Ray Image Detectors for Digital Fluoroscopy and Fluorography, Clin. Radiol. 63 (2008) 1073−1085.

[10] H. Wieczorek, O. Overdick, Afterglow and Hysteresis in CsI:Tl. In Proceedings of the 5th International Conference on Inorganic Scintillators and Their Applications (SCINT99), Moscow, Russia, August 16−20 1999; Mikhailin, V., Ed.; M.V. Lomonosov Moscow State University: Moscow, Russia, 2000, pp 385−390.

[11] V.V. Nagarkar, T.K. Gupta, S.R. Miller, Y. Klugerman, M.R Squillante, G. Entine, Structured CsI(Tl) Scintillators for X-ray Imaging Applications, IEEE Trans. Nucl. Sci. 45 (1998) 492−496.

[12] J. Yorkston, Recent Developments in Digital Radiography Detectors, Nucl. Instrum. Methods Phys. Res., Sect. A 580 (2007) 974−985.

[13] E. Dell'Orto, M. Fasoli, G. Ren, A. Vedda, Defect-Driven Radioluminescence Sensitization in Scintillators: The Case of $Lu_2Si_2O_7$:Pr, J. Phys. Chem. C 117 (20130 20201−20208.

[14] M. Buryi, V. Laguta, M. Nikl, V. Gorbenko, T. Zorenko, Yu. Zorenko, LPE growth and study of the $Ce^{3+}$ incorporation in $LuAlO_3$:Ce single crystalline film scintillators, Cryst. Eng. Comm. 21 (2019) 3313-3321.

[15] D.A. Spassky, N.S. Kozlova, A.P. Kozlova, E.V. Zabelina, O.A. Buzanov, M.Buryi, V. Laguta, K. Lebbou, A. Nehari, H. Cabane, M. Dumortier, V. Nagirnyi, Study of the defects in $La_3Ta_{0.5}Ga_{5.5}O_{14}$ single crystals, Journal of Luminescence 180 (2016) 95–102.

[16] V. Jarý, L. Havlák, J. Bárta, M. Buryi, E. Mihóková, M. Rejman, V. Laguta, M. Nikl, Optical, Structural and Paramagnetic Properties of Eu-Doped Ternary Sulfides $ALnS_2$ (A = Na, K, Rb; Ln = La, Gd, Lu, Y), Materials 8 (2015) 6978–6998.

[17] M. Nikl, V.V. Laguta, A. Vedda, Complex oxide scintillators: Material defects and scintillation performance, Phys. Stat. Sol. (b) 245 (2008) 1701–1722.

[18] M. Buryi, P. Bohacek, K. Chernenko, A. Krasnikov, V.V. Laguta, E. Mihokova, M. Nikl, S. Zazubovich, Luminescence and photo-thermally stimulated defect-creation processes in $Bi^{3+}$-doped single crystals of lead tungstate. Phys. Stat. Solidi B 253, (2016) 895–910.

[19] V.V. Laguta, A.M. Slipenyuk, M.D. Glinchuk, I.P. Bykov, Y. Zorenko, M. Nikl, J. Rosa, K. Nejezchleb, Paramagnetic impurity defects in LuAG:Ce thick film scintillators, Radiation Measurements 42 (2007) 835–838.

[20] R.A. Serway, F.H. Yang, S.A. Marshall, Line Broadening in the ESR Absorption Spectra of $Fe^{3+}$, $Cr^{3+}$, and $Gd^{3+}$ in the Oxide YAlG, Phys. Stat. Solidi (b) 89 (1978) 267.

[21] Kh. Bagdasarov, V.V. Bershov, V.O. Martirosyan, M.I. Meilman, The State of Molybdenum Impurity in Yttrium-Aluminium Garnet, Phys. Stat. Solidi (b) 46 (1971) 745.

[22] R. Jablonski, S.M. Kaczmarek, Electron Spin Resonance and Optical measurements in YAG:$V^{3+}$ crystals, Proc. SPIE 3724, International Conference on Solid State Crystals '98: Single Crystal Growth, Characterization, and Applications, 1999, 34-6352.

[23] G.M. Zverev, A.M. Prokhorov, Electron paramagnetic resonance spectrum of $V^{3+}$ in corundum, Sov. Phys. - JETP 11 (1960) 330–333.

[24] J. Isoya, J.A. Weil, R.F.C. Claridge, The dynamic interchange and relationship between germanium centers in a-quartz, J. Chem. Phys. 69 (1978) 4876.

[25] D.L. Griscom, E.J. Friebele, K.J. Long, J.W. Fleming, Fundamental defect centers in glass: Electron spin resonance and optical absorption studies of irradiated phosphorus-doped silica glass and optical fibers, J. Appl. Phys. 54 (1983) 3743-3762.

[26] V.V. Laguta, M. Buryi, J. Rosa, D. Savchenko, J. Hybler, M. Nikl, S. Zazubovich, T. Karner, C.R. Stanek, and K.J. McClellan, Electron and hole traps in yttrium orthosilicate single crystals: The critical role of Si-unbound oxygen, Phys. Rev. B 90 (2014) 064104.

[27] E. Dell'Orto, M. Fasoli, G. Ren, A. Vedda, Defect-Driven Radioluminescence Sensitization in Scintillators: The Case of $Lu_2Si_2O_7$:Pr, J. Phys. Chem. C 117 (2013) 20201−20208.

[28] E. Mihóková, M. Fasoli, F. Moretti, M. Nikl, V. Jary, G. Ren, A. Vedda, Defect states in $Pr^{3+}$ doped lutetium pyrosilicate, Optical Materials 34 (2012) 872–877.

[29] L. Pidol, O. Guillot-Noel, M. Jourdier, A. Kahn-Harari, B. Ferrand, P. Dorenbos, D. Gourier, Scintillation quenching by $Ir^{3+}$ impurity in cerium doped lutetium pyrosilicate crystals, J. Phys.: Condens. Matter. 15 (2003) 8715-7821.

[30] F. Bretheau-Raynal, M. Lance, P. Charpin, Crystal data for $Lu_2Si_2O_7$, J. Appl. Cryst. 14 (1981) 349-350.

[31] C.P Poole, Jr. and H.A. Farach, Eds. Handbook of Electron Spin Resonance, Springer-Verlag, Inc.: New York, NY, USA, 1999, Volume 2.

[32] G.H. Fuller, Nuclear spins and moments, J. Phys. Chem. Ref. Data 5 (1976) 835.

[33] V. Laguta, M. Buryi, A. Beitlerova, O. Laguta, K. Nejezchleb, M. Nikl, Vanadium in yttrium aluminum garnet: charge states and localization in the lattice, Opt. Materials 91 (2019) 228–234.

[34] A. Abragam, B. Bleaney, Electron Paramagnetic Resonance of Transition Ions, Clarendon Press, Oxford, 1970.

[35] P. Neugebauer, D. Bloos, R. Marx, P. Lutz, M. Kern, D. Aguila, J. Vaverka, O. Laguta, C. Dietrich, R. Clerac, J. van Slageren, Ultra-broadband EPR spectroscopy in field and frequency domains, Phys. Chem. Chem. Phys. 20 (2018) 15528.

[36] R.D. Shannon, Revised effective ionic radii and systematic studies of interatomic distances in halides and chalcogenides, Acta Crystallogr. A 32 (1976) 751–767.

[37] H. Feng, G. Ren, Y. Wu, J. Xu, Q. Yang, J. Xie, M. Chou, C. Chen, Optical and thermoluminescence properties of $Lu_2Si_2O_7$:Pr single crystal, Journal of rare earths 30 (2012) 775-779.

[38] L. Pidol, B. Viana, A. Kahn-Harari, A. Bessiere, P. Dorenbos, Luminescence properties and scintillation mechanisms of $Ce^{3+}$-, $Pr^{3+}$- and $Nd^{3+}$-doped lutetium pyrosilicate, Nuclear Instruments and Methods in Physics Research A 537 (2005) 125–129.

[39] M. Buryi, V.V. Laguta, E. Mihokova, P. Novak, M. Nikl, Electron paramagnetic resonance study of the $Ce^{3+}$ pair centers in $YAlO_3$:Ce scintillator crystals, Phys. Rev. B 92 (2015) 224105.

[40] L. Zheng, G. Zhao, C. Yan, G. Yao, X. Xu, L. Su, J. Xu, Growth and spectroscopic characteristics of Yb:LPS single crystal, Journal of Crystal Growth 304 (2007) 441–447.